# 3D printing of a leaf spring: A demonstration of closed-loop control in additive manufacturing

Kévin Garanger[1], Thanakorn Khamvilai[2], and Eric Feron[3]

*Abstract*— This paper presents the integration of a feedback control loop during the printing of a plastic object using additive manufacturing. The printed object is a leaf spring made of several parts of different infill density values, which are the control variables in this problem. In order to achieve a desired objective stiffness, measurements are taken after each part is completed and the infill density is adjusted accordingly in a closed-loop framework. The absolute error in the stiffness at the end of printing is reduced from $11.63\%$ to $1.34\%$ by using a closed-loop instead of an open-loop control. This experiments serves as a proof of concept to show the relevance of using feedback control in additive manufacturing. By considering the printing process and the measurements as stochastic processes, we show how stochastic optimal control and Kalman filtering can be used to improve the quality of objects manufactured with rudimentary printers.

## I. INTRODUCTION

Additive manufacturing technologies have become increasingly popular for the production of complex parts when other traditional methods cannot be used or require the manufacturing of large batches to be economically viable. However, the high variability in the quality of the builds printed using additive manufacturing (AM) is an obstacle that limits the impact that AM can have in sensitive applications [1]. The introduction of closed-loop control in AM is a main stake of research in this domain because it would allow better reliability guarantees for the objects being built [2]–[4]. But the high number of control variables makes it difficult to completely understand their impact on the relevant properties of the final build (for instance mechanical or thermal). The important design features often refer to global properties that are directly induced by the microscopic material properties, but are hard to relate to the control inputs. Deriving meaningful relationships between control inputs and final properties is a challenge of AM, and a task necessary to the definition of efficient feedback control laws. Another requirement for introducing real-time feedback control in AM is to have precise measurements during the printing process and to relate them with the expected final properties of the build [5]. Several works developed systems with closed-loop controls capabilities such as [6] or [7]. Those methods, however, focus on the control of microscopic variables without taking in account the final properties of the object being built. In the recent publication [8], the authors design a closed-loop control system that detects and corrects filament bonding failures for fused deposition modeling (FDM). The experiment performed in our work is a demonstration of closed-loop control in additive manufacturing. We consider the printing of a leaf spring using FDM, and we are interested in taking intermediate measurements after the completion of each part of the leaf spring to estimate the stiffness along one axis of the build and reconsider the infill density. Objects printed with AM often use porous structures with different possible patterns because of the good mechanical properties and the gain of weight obtained by those structures [9]–[11]. The density of such a pattern is the control variable modified throughout this experiment. The dynamics of the system are not based on a physical model relating the input (the infill density) with the output (the stiffness of the printed object), but are based on a purely statistical model. Preliminary measurements were performed on test specimens to show that such a model can be used, and to determine the parameters of this model. Our goal is to show how feedback control through *in situ* measurements can be beneficial to the field of AM. A larger objective is to incite to the development of a broad framework to characterize the properties of objects printed with AM and to use them for deriving feedback control laws. We first describe in details the setting of the experiment performed in section II. Then, from a basic probabilistic model relating the input and output of our system, we derive an optimal control law in section III. Finally, in section IV, we describe the results of the experiment before giving some concluding remarks in section V.

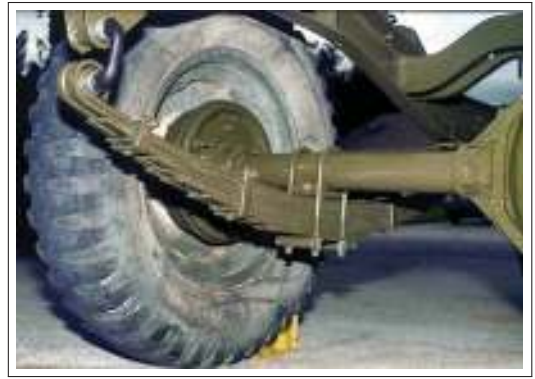

Fig. 1: Picture of a leaf spring of a Willys M38 [12]

## II. ADDITIVE MANUFACTURING OF A LEAF SPRING

### A. Process description

This experiment consists in the additive manufacturing of a leaf spring. Leaf springs are springs made of several

[1,2,3]All authors are with the Department of Aerospace Engineering at Georgia Institute of Technology, Atlanta, GA, 30332, USA {kevin.garanger, tkhamvilai3, feron}@gatech.edu.

stacked leaves that are commonly used for the suspensions of wheeled vehicles [13] (Figure 1). Because a leaf spring is made of several parts built independently and then assembled, its manufacturing is a sequential problem that fits perfectly the framework of a discrete dynamic programming problem. Each step corresponds to the printing of a new leaf and the applied control is chosen to reach a final objective. In this case, a stack of $n$ leaves is designed to have a fixed final geometry and a specific stiffness along the vertical axis (Figure 2). The stiffness of a leaf is defined a the linear coefficient relating deflection to load applied during a 3-point bending test, assuming a linear relationship. Each leaf is made of the same number of layers and of the same material. Leaves are assumed to be Euler-Bernoulli beams [14] and a 3-point bending test is used to measure their stiffness. In order to achieve the desired stiffness objective, the infill density of each new leaf is adapted in a closed-loop setting. To do so, measurements are performed after the printing of each leaf to evaluate the stiffness of the partially built object and to meet a target overall stiffness. Because the different leaves of the leaf spring are not stuck together, the stiffness of a stack of leaves is approximated as additive in the Euler-Bernoulli theory. Leaves are printed independently before stacking them to ensure that this condition is respected. The additivity property of the stiffness allows the use of a linear Kalman filter to estimate the stiffness at each step more precisely. The derivation of the filtering that is used is detailed in section III while the parameters of the filters are estimated with some preliminary measurements. The results of those measurements are given in section IV.

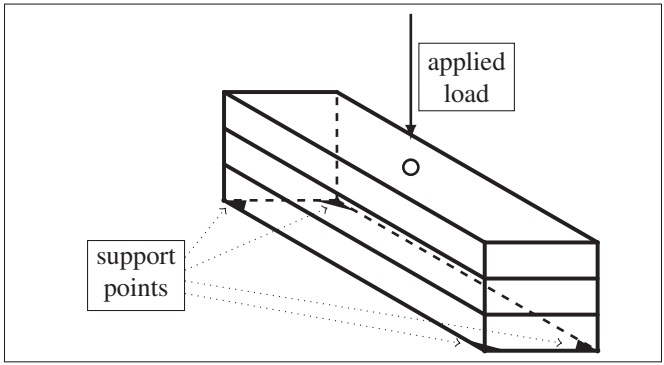

Fig. 2: Stack of 3 leaves with a load applied on top and 4 supports on the bottom corners for the stiffness measurement

### B. Experimental setting

*1) Printing procedure:* For this experiment, a low-cost printer was chosen since the objective of this work is proving that feedback control based on in situ measurements can be used to print more reliably with material subject to a high process noise. The Printrbot Simple 3D printer - 1405 Model [15] (Figure 3) is chosen because more random variation is expected during the printing process from such a printer than with a high-performance one [16]. The filament type used in this experiment is Polylactic Acid (PLA), which is provided with the printer package. The identical Computer-Aided-Design (CAD) model of every specimen is developed using Solidworks [17]. The G-Code [18] files are generated using the default setting of Cura [19], except for the percentage of infill density. Finally, Pronterface [20] is used as a graphic user interface (GUI) for monitoring and communicating between the 3D printer and a computer. Some example specimens with different percentages of infill density are showed in Figure 4.

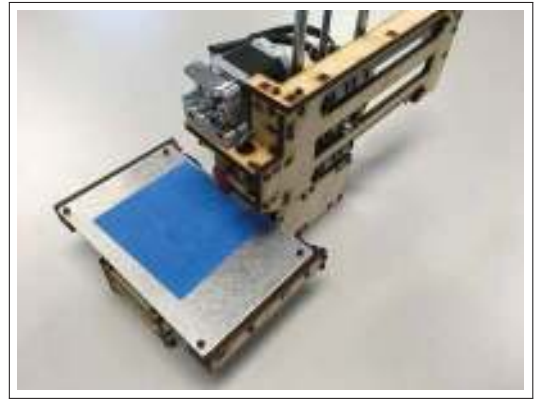

Fig. 3: Printrbot Simple 3D printer - 1405 Model

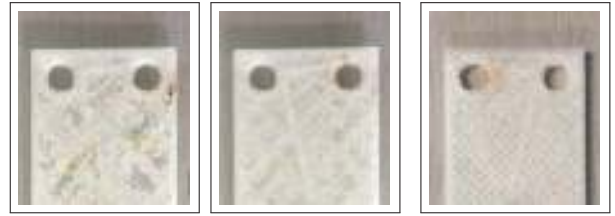

(a) 10% infill density    (b) 20% infill density    (c) 30% infill density

Fig. 4: Example specimens with different percentages of infill density

*2) Stiffness measurement procedure:* In this experiment, the preparation of the PLA specimens and of the three-point bending test (Figure 5) is performed based on ASTM D790 [21], which is the standard testing method for flexural properties of plastic materials. Since each specimen is required to be stacked over the next one, we constrained our experiment within the elastic region of the material. Then a load acting on the specimen and its vertical deflection were measured at each time step. After that, the stiffness is determined from the slope of the linear regression between the deflection and load data sets (Figure 6). Note that in this figure, some geometric nonlinearities can be observed, suggesting that the tests performed were not restricted to the domain of elasticity of the specimens. A better model would require more careful measurements. However, the objective of this work is not exactly to derive a precise model but rather to show that closed-loop control can be relevant without a perfect model.

### III. FEEDBACK CONTROL LAW

In this section, an optimal control law that aims at reaching a target stiffness while minimizing a specified cost function is derived. At each step $i$, a new measurement of the stiffness is performed and taken into account to refine the estimate of the predicted stiffness at the final step $n$. This is done by using filtering to estimate the actual stiffness of a stack of leaves. Two types of noises are considered: a process noise that comes from the inaccuracy of the printer and from the changing environment, and a measurement noise. Both are assumed to follow independent normal laws. In the following subsections the process to estimate the stiffness of a stack at each step is described. Then this stiffness estimate is used to obtain an optimal control law. The chosen parameterization of the process noise is also detailed while the algorithm obtained from the optimal control law is described.

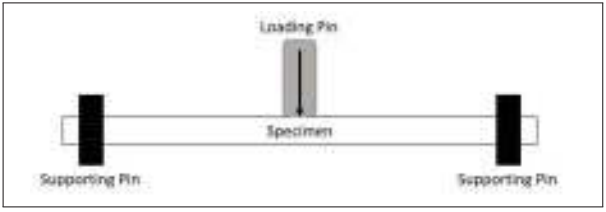

Fig. 5: The setup of the 3-point bending test

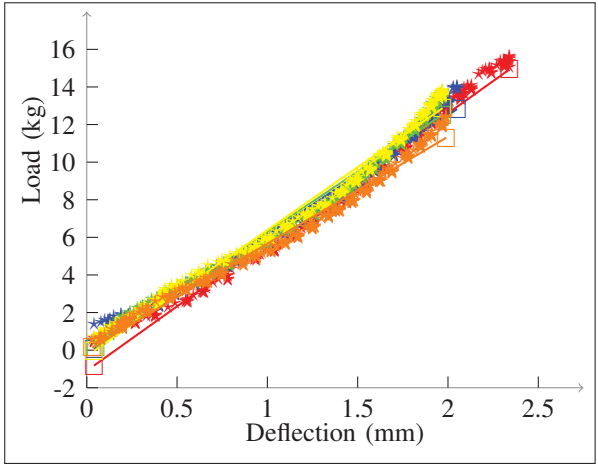

Fig. 6: Plot between an applied load and a vertical deflection of a single specimen with 10% infill density from 5 measurements. Each color represents a measurement made of several data points represented by ⋆. Lines represent the linear regressions of these data points.

### A. Estimating the stiffness of a stack of leaves

In this subsection the stiffness of a stack of printed leaves is estimated given the controls that have been previously applied and given the measurements after each new printed leaf. This is equivalent to applying a linear Kalman filter. For a sequence of controls $(d_i)_{i \leq n}$, let $K_i$ be the stiffness of the first $i$ printed leaves stacked together. This is a random variable defined recursively by

$$K_0 = 0$$

and

$$K_{i+1} = K_i + \mu_p(d_{i+1}) + \epsilon_{i+1} \qquad (1)$$

where

$$\epsilon_i \sim \mathcal{N}(0, \sigma_p)$$

and

$$\mu_p(d_i)$$

are respectively independent identically distributed random variables and the mean stiffness of a single leaf of density $d_i$.

The stiffness observations of a stack of the first $i$ leaves are also defined by

$$\bar{K}_i = K_i + \bar{\epsilon}_i \qquad (2)$$

where

$$\bar{\epsilon}_i \sim \mathcal{N}(0, \sigma_o)$$

are independent identically distributed random variables independent of each process noise $(\epsilon_j)_{j \leq n}$, previous stiffnesses observations $(\bar{K}_j)_{j < i}$, and past controls $(d_j)_{j \leq i}$.

To derive the probability law of the stiffness $K_i$ given the previous observations and past controls, the Bayes rule is applied to the joint probability of $K_i$ and $\bar{K}_i$.

$$p(K_i|(\bar{K}_j)_{j \leq i}, (d_j)_{j \leq i}) p(\bar{K}_i|(\bar{K}_j)_{j < i}, (d_j)_{j \leq i})$$
$$= p(K_i, \bar{K}_i|(\bar{K}_j)_{j < i}, (d_j)_{j \leq i})$$
$$= p(\bar{K}_i|K_i, (\bar{K}_j)_{j < i}, (d_j)_{j \leq i}) p(K_i|(\bar{K}_j)_{j < i}, (d_j)_{j \leq i})$$

Since the stiffness observations are independent, it yields

$$p(K_i|(\bar{K}_j)_{j \leq i}, (d_j)_{j \leq i}) p(\bar{K}_i|(\bar{K}_j)_{j < i}, (d_j)_{j \leq i})$$
$$= p(\bar{K}_i|K_i) p(K_i|(\bar{K}_j)_{j < i}, (d_j)_{j \leq i}) \qquad (3)$$

With the observation relation (2), the probability distribution of the observation is given by

$$p(\bar{K}_i|K_i) = \frac{1}{\sqrt{2\pi\sigma_o^2}} \exp\left(-\frac{1}{2}\frac{(\bar{K}_i - K_i)^2}{\sigma_o^2}\right)$$

Then, using the process relation (1),

$$p(K_i|(\bar{K}_j)_{j < i}, (d_j)_{j \leq i})$$
$$= \int_{K_{i-1}} p(K_i|K_{i-1}, d_i) p(K_{i-1}|(\bar{K}_j)_{j < i}, (d_j)_{j \leq i}) \mathrm{d}K_{i-1}$$
$$= \int_{K_{i-1}} \frac{1}{\sqrt{2\pi\sigma_p^2}} \exp\left(-\frac{1}{2}\frac{(K_i - K_{i-1} - \mu_p(d_i))^2}{\sigma_p^2}\right)$$
$$\times p(K_{i-1}|(\bar{K}_j)_{j < i}, (d_j)_{j < i}) \mathrm{d}K_{i-1} \qquad (4)$$

Reusing equation (2) leads to

$$p(\bar{K}_i|(\bar{K}_j)_{j < i}, (d_j)_{j \leq i})$$
$$= \int_{K_i} p(\bar{K}_i|K_i) p(K_i|(\bar{K}_j)_{j < i}, (d_j)_{j \leq i}) \mathrm{d}K_i$$
$$= \int_{K_i} \frac{1}{\sqrt{2\pi\sigma_o^2}} \exp\left(-\frac{1}{2}\frac{(\bar{K}_i - K_i)^2}{\sigma_o^2}\right)$$
$$\times p(K_i|(\bar{K}_j)_{j < i}, (d_j)_{j \leq i}) \mathrm{d}K_i \qquad (5)$$

Therefore, these three equations give a recursive relation for the derivation of $p(K_i|(\bar{K}_j)_{j \leq i}, (d_j)_{j \leq i})$.

Assuming that $p(K_{i-1}|(\bar{K}_j)_{j<i},(d_j)_{j<i})$ is the probability of a normal law of mean $\mu_{i-1}$ and variance $\sigma_{i-1}$, replacing in equation (4) gives

$$p(K_i|(\bar{K}_j)_{j<i},(d_j)_{j\leq i}) = \frac{1}{\sqrt{2\pi(\sigma_p^2 + \sigma_{i-1}^2)}} \exp\left(-\frac{1}{2}\frac{(K_i - \mu_{i-1} - \mu_p(d_i))^2}{\sigma_p^2 + \sigma_{i-1}^2}\right)$$

Then, replacing in equation (5),

$$p(\bar{K}_i|(\bar{K}_j)_{j<i},(d_j)_{j\leq i}) = \frac{1}{\sqrt{2\pi(\sigma_o^2 + \sigma_p^2 + \sigma_{i-1}^2)}} \exp\left(-\frac{1}{2}\frac{(\bar{K}_i - \mu_{i-1} - \mu_p(d_i))^2}{\sigma_o^2 + \sigma_p^2 + \sigma_{i-1}^2}\right)$$

These two results are plugged into the observation law in equation (3), which gives

$$p(K_i|(\bar{K}_j)_{j\leq i},(d_j)_{j\leq i}) = \frac{1}{\sqrt{2\pi\sigma_i^2}} \exp\left(-\frac{1}{2}\frac{(K_i - \mu_i)^2}{\sigma_i^2}\right)$$

With

$$\mu_i = \frac{\bar{K}_i(\sigma_p^2 + \sigma_{i-1}^2) + (\mu_{i-1} + \mu_p(d_i))\sigma_o^2}{\sigma_o^2 + \sigma_p^2 + \sigma_{i-1}^2} \quad (6)$$

and

$$\sigma_i^2 = \frac{\sigma_o^2(\sigma_p^2 + \sigma_{i-1}^2)}{\sigma_o^2 + \sigma_p^2 + \sigma_{i-1}^2} \quad (7)$$

By initializing with $\mu_0 = 0$ and $\sigma_0 = 0$, it is easy to verify that the conditional probability of $K_1$ given $\bar{K}_1$ and $d_1$ indeed follows a normal law with parameters given by equations (6) and (7) with $i = 1$. An induction argument proves that those relations are true for every $i \geq 1$.

*Remark 1:* Equation (7) defines a Riccati difference equation. It can be solved to provide an expression of $\sigma_i^2$ independent of $\sigma_{i-1}^2$.

*Remark 2:* The process noise variance $\sigma_p^2$ is assumed to be a constant variable but the results would be unchanged if it depended on the control $d_i$.

*Remark 3:* The observation noise variance $\sigma_o^2$ can be reduced by taking several measurements of the same stack. In that case in equation (7) the observation noise variance will simply be divided by the number of observations.

### B. Estimating the final stiffness

Let $\mu_i$ be the stiffness of the first $i$ stacked leaves taken altogether. Given the next controls $(d_j)_{i<j\leq n}$ too, the final stiffness $K_n$ of the stacked $n$ leaves can be estimated by:

$$\mathbb{E}(K_n|(\bar{K}_j)_{1\leq j\leq i},(d_j)_{1\leq j\leq n}) = \mu_i + \sum_{j=i+1}^{n} \mu_p(d_j)$$

### C. Optimal control of the printing process

In this section, the controls $(d_j)_{j\leq n}$ are derived by minimizing the expectation of a cost function $J(d_1,\ldots,d_n,K_1,\ldots,K_n)$ while reaching the objective stiffness $K$.

At step $i$, let $(d_j^*)_{j\leq i}$ be the chosen values at the previous steps. Let $\mathcal{H}_i$ be the set of real-valued $(d_j)_{i<j\leq n}$ verifying the equation

$$\sum_{j=i+1}^{n} \mu_p(d_j) = K - \mu_i$$

With this definition, the next controls are $n - i$ values $(d_j^i)_{i<j\leq n}$ such that

$$\mathbb{E}\big(J(d_1^*,\ldots,d_i^*,d_{i+1}^i,\ldots,d_n^i,K_1,\ldots,K_n) \big| (\bar{K}_j)_{1\leq j\leq i},(d_j^*)_{1\leq j\leq i},(d_j^i)_{i<j\leq n}\big)$$
$$= \min_{(d_j)\in\mathcal{H}_i} \mathbb{E}\big(J(d_1^*,\ldots,d_i^*,d_{i+1},\ldots,d_n,K_1,\ldots,K_n) \big| (\bar{K}_j)_{1\leq j\leq i},(d_j^*)_{1\leq j\leq i},(d_j)_{i<j\leq n}\big)$$

And then the control $d_{i+1}^i$ is applied, such that $d_{i+1}^* := d_{i+1}^i$.

In the following subsection two examples of cost functions are given.

*1) Minimizing the quantity of used material:* A possible cost function characterizes the quantity of used printing material:

$$J = \sum_{j=1}^{n} d_j$$

However because the relationship between density and stiffness is linear, the cost function has same value everywhere on $\mathcal{H}_i$, leading to infinitely many possibilities. Instead, the sum of the squares of the densities can be used:

$$J = \sum_{j=1}^{n} d_j^2$$

In that case, because of the symmetric roles of the different remaining leaves in the cost function and in the constraints, all $(d_j^i)_{i<j\leq n}$ are equal and

$$\mu_p(d_j^i) = \frac{1}{n-i}(K - \mu_i), \forall j, i < j \leq n \quad (8)$$

This is the cost function that is used in the rest of the experiment.

### D. Process noise parameterization

In the experiments performed in the next section, the mean stiffness is assumed to be affine in $d$ and the process variance is assumed to be constant.

$$\mu_p(d) = \alpha d + \beta$$

and

$$\sigma_p(d) = \sigma_p > 0$$

These assumptions are based on previous performed measurements on different leaf specimens. Results of these measurements are detailed in section IV.

With these assumptions, at each step the optimal control is given by

$$d_{i+1}^* = \frac{K - \mu_i}{\alpha(n - i)} - \frac{\beta}{\alpha}, \forall k, i < k \leq n \quad (9)$$

### E. Filtering algorithm

The derivation of an optimal input density at each step based on an estimate of the stiffness using filtering is the basis of algorithm 1.

---

**Algorithm 1** Optimal printing algorithm with filtering

---

**Require:** $n, K, \alpha, \beta, \sigma_p, \sigma_o$
  *Initialization:*
1: $\mu = 0$
2: $\sigma^2 = 0$
  *Printing:*
3: **for** $i = 0$ to $n - 1$ **do**
4: $\quad d := \frac{K - \mu}{\alpha(n-i)} - \frac{\beta}{\alpha}$
5: $\quad$ Print a leaf with input density $d$
6: $\quad$ Measure stiffness of the printed leaves $\bar{K}$
7: $\quad$ Update $\mu := \frac{\bar{K}(\sigma_p^2 + \sigma^2) + (\mu + \mu_p(d_i))\sigma_o^2}{\sigma_o^2 + \sigma_p^2 + \sigma^2}$
8: $\quad$ Update $\sigma^2 := \frac{\sigma_o^2(\sigma_p^2 + \sigma^2)}{\sigma_o^2 + \sigma_p^2 + \sigma^2}$
9: **end for**

---

Besides specifying the number of stacks $n$ and the desired stiffness $K$, algorithm 1 requires the knowledge of the density-stiffness affine model parameters $\alpha$ and $\beta$, the process noise standard deviation $\sigma_p$, and the observation noise standard deviation $\sigma_o$, which all can be obtained from the prior measurement data.

After initialization, there are two essential steps during the printing. The first one is picking an optimal infill density $d_i^*$ for the leaf to print in line 4 based on the estimate of the stiffness and on the density-stiffness model. This step is the control determination step.

The second step which is the measurement update of line 7 and 8, updates the intermediate parameters $\mu$ and $\sigma$ with the measured stiffness $\bar{K}$.

The process of this algorithm is summarized by the block-diagram in Figure 7.

## IV. EXPERIMENTS

In this section, two different sets of experiments are described. The first set of experiments aimed at validating the hypothesis that the process law mean is affine and that the variance of both process and observation noise can be assumed constant. The second set of experiments consists in printing a stack of leaves using the filtering algorithm for which the obtained results are presented. These results are compared to the ones obtained with two different baselines. The first baseline consists in printing the stack of leaves without any feedback control (open-loop). In that case all leaves have the density that is determined before printing and no measurement is performed during the printing process. For the second baseline, no filtering is used (closed-loop without filtering). The stiffness used to determine the next density input at each step is the value of the measurement at that step. This is equivalent to considering that there is no observation noise.

### A. Determination of the process and observation noises

To evaluate the parameters of the process noise and of the observation noise, a set of 15 single leaves with 5 different input densities was printed. A three-point bending test was performed on each of them and a dataset of loads vs. deflection was acquired. Using a linear regression, the stiffness measured stiffness was obtained. 5 set measurements per leaf were performed Results are presented in Tables II, III, IV of the appendix, and final stiffnesses are showed in Figure 8. A first-order regression was then performed between the measured stiffnesses and the infill densities to obtain the values of $\alpha$ and $\beta$. The process noise was determined by taking the standard deviation of the means of the measurements per specimen, whereas the observation noise was determined by taking the mean of the standard deviations of the measurements per specimen. With these results the following parameterizations for the process and observation noises are found:

$$\mu_p(d) = 0.3073d + 4.5593 \quad (10)$$
$$\sigma_p = 1.0579$$
$$\sigma_o = 0.6907$$

### B. Printing the leaf spring

Leaf springs were printed for different values of $n$ and $K$ under the three previously described methods. The two combinations tried for the pair $(n, K)$ are $(3, 30)$ and $(3, 40)$. When performing stiffness measurements, the mean of 5 subsequent measurements was taken. The results of the stiffness measurements are reported in Table I, Figure 9, and 10. As shown in these figures, the filtering leads to a final stiffness closer to the objective than the baseline methods do.

Consider the case $n = 3, K = 30$ (Figure 9), at the first step every process starts with the specimen of the same density, which is the best value according to the prior knowledge. Once the measurement has been performed for the closed-loop processes, both of them have the nearly-identical value of stiffness due to the small observation noise. Nevertheless, this value is not exactly the desired one. At the next step, the feedback control corrects that error from the previous step. However, the controller performs better when the stiffness is estimated using the forgoing filter. In the non-filtering case, the stiffness measurement is considered perfect and the information of the control that led to that stiffness is discarded. At the final step, the closed-loop control with filtering reached a better stiffness than both baselines.

Similarly for the case $n = 3, K = 40$ (Figure 10), the closed-loop control with filtering provides a better result, even though in the first two steps, the non-filtering controller has its measured values closer to the nominal one.

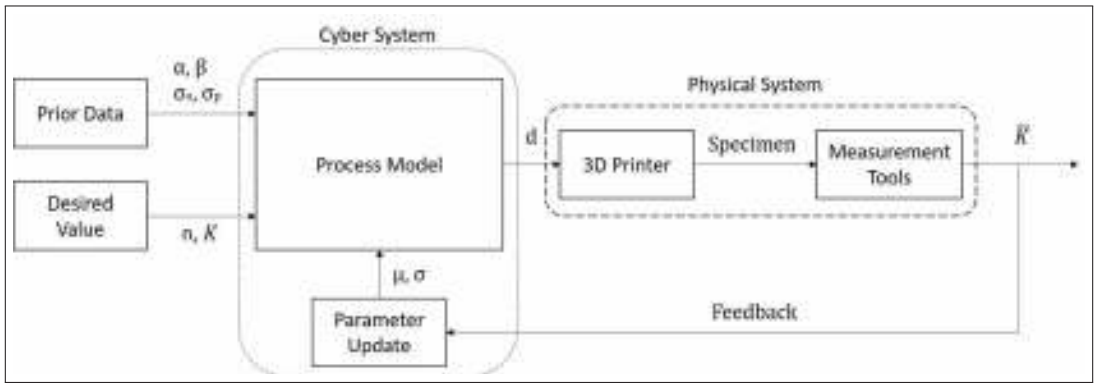

Fig. 7: The process block-diagram

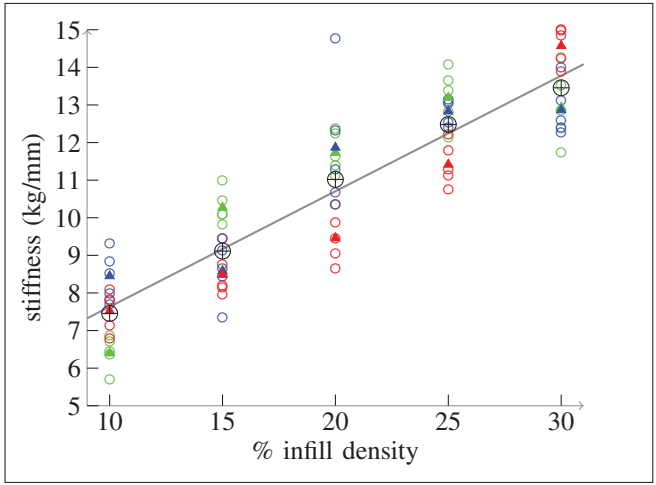

Fig. 8: Stiffness measurements for 15 specimens of different infill densities. ○ represents each measurement. △ represents the average stiffness of each specimen from 5 measurements. ⊕ represents the average stiffness of specimens with the same infill density. R, G, and B represent the $1^{st}$, $2^{nd}$, and $3^{rd}$ specimen of the same infill density, respectively. The line represents the linear regression of ⊕ data given by equation (10).

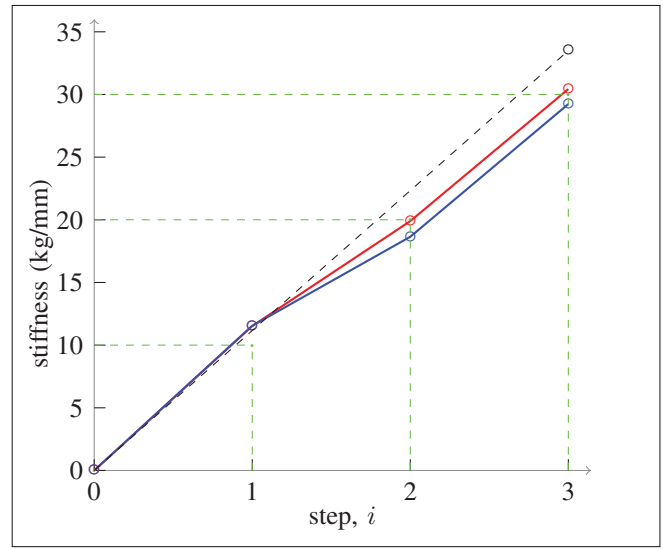

Fig. 9: Stiffness measurements at steps $1, 2, 3$ for $n = 3, K = 30$. ○ represents a measurement. Red is used for closed-loop with filtering. Blue is used for closed-loop without filtering. Dashed **black** is used for open-loop. Green marked coordinates represent the desired values of stiffness.

TABLE I
Measured stiffness (kg/mm) at steps 1, 2, 3 using the filtering algorithm and the two baselines with the absolute error in the stiffness at each final step

| $(n, K)$ | Closed-loop with filtering | | | Closed-loop without filtering | | | Open-loop | | |
|---|---|---|---|---|---|---|---|---|---|
| (3, 30) | 11.53 | 19.89 | **30.43** | 11.55 | 18.65 | **29.24** | – | – | **33.49** |
| Error | 1.43 % | | | 2.53 % | | | 11.63 % | | |
| (3, 40) | 12.53 | 27.86 | **40.89** | 12.67 | 26.31 | **42.29** | – | – | **37.09** |
| Error | 2.23 % | | | 5.73 % | | | 7.28 % | | |

## V. CONCLUSION

In this paper, we presented the idea and implementation of a feedback control system for a specific additive manufacturing process. The feedback control is based on measurements taken during the process and aims at reaching a specific desired stiffness for an object comparable to a leaf spring. A better precision was achieved using a closed-loop control with filtering than by using two baselines: a closed-loop control without filtering and an open-loop control. This experiment, while very specific and hardly generalizable as it is, shows the relevance of feedback control in AM. One may argue that what makes this experiment successful is not directly related to the fact that AM is used for manufacturing the final part. But even if a similar experiment could certainly be imagined without using a 3D printer but with another process instead, the material requirements would have been limiting and the whole process more complicated and time-consuming. The field of AM is appealing because of the simplicity it offers for manufacturing complex and various parts quickly. And we think that it could benefit from the introduction of closed-loop control systems based on widespread methods coming from subfields of optimal control such as stochastic optimal control and filtering. The use of dynamic programming and better control algorithms in general, based on the expected characteristics of an object and nondestructive testing, could leverage the potential of low-cost printers to produce high-quality prints. To generalize this experiment to different processes and desired object properties, a general framework to describe various object properties and link them to control variables is required. Moreover, nondestructive testing technologies such as computed tomography scanning or ultrasonic testing have to be better integrated with 3D printers to perform *in-situ* measurements during printing.

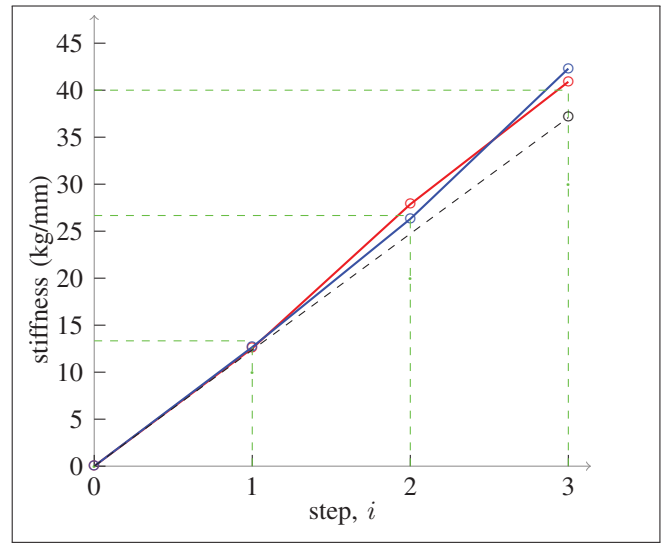

Fig. 10: Stiffness measurements at steps $1, 2, 3$ for $n = 3, K = 40$. Legend is same as in Fig. 9

## APPENDIX

The experimental results which are mean stiffness, standard derivation of stiffness, and the percentage of infill density of specimen are given in tables II, III, IV, and V.

TABLE II
Mean of the stiffness of each specimen from 5 measurements (kg/mm)

| density | 1st specimen | 2nd specimen | 3rd specimen |
|---|---|---|---|
| 10% | 6.4024 | 7.5183 | 8.4502 |
| 15% | 10.2724 | 8.4851 | 8.5846 |
| 20% | 11.7310 | 9.4587 | 11.8682 |
| 25% | 13.1919 | 11.4165 | 12.8316 |
| 30% | 12.9317 | 14.5737 | 12.8644 |

TABLE III
Standard deviation of the stiffness of each specimen from 5 measurements (kg/mm)

| density | 1st specimen | 2nd specimen | 3rd specimen |
|---|---|---|---|
| 10% | 0.4432 | 0.5480 | 0.6473 |
| 15% | 0.4524 | 0.5990 | 0.8108 |
| 20% | 0.5414 | 0.6596 | 1.7784 |
| 25% | 0.7581 | 0.5805 | 0.3430 |
| 30% | 0.9821 | 0.5072 | 0.7098 |

TABLE IV
Mean and standard deviation of the stiffness of specimens of equal density (kg/mm)

| density | mean | standard deviation |
|---|---|---|
| 10% | 7.4570 | 1.0253 |
| 15% | 9.1140 | 1.0044 |
| 20% | 11.0193 | 1.3533 |
| 25% | 12.4800 | 0.9385 |
| 30% | 13.4566 | 0.9680 |

TABLE V
% Printed infill density of the specimen at steps 1, 2, 3 using the filtering algorithm and the two baselines obtained from equation (9)

| $(n, K)$ | Closed-loop with filtering | | | Closed-loop without filtering | | | Open-loop (at every step) |
|---|---|---|---|---|---|---|---|
| (3, 30) | 17.705 | 15.475 | 17.375 | 17.705 | 15.186 | 22.108 | 17.705 |
| (3, 40) | 28.552 | 29.648 | 24.808 | 28.552 | 29.634 | 29.734 | 28.552 |

## ACKNOWLEDGMENT

We would like to thank Juan Pablo Afman for his help and suggestions regarding the design and manufacturing of the specimens and the stiffness testing process.